\documentclass[11pt, a4paper]{article}
\usepackage{jcappub, physunits, slashed}
\usepackage{setspace, caption, subcaption}
\usepackage{nccmath, physics, tensor}
\usepackage{xparse, mathtools, aas_macros}
\usepackage[toc, page]{appendix}
\usepackage[mathscr]{euscript}
\usepackage{tikz}
\usepackage[compat=1.0.0]{tikz-feynman}
\usepackage{tikzfeynman}

\usepackage{graphicx}

\title{Reheating with Axion-SU(2) and Gravitational Chern-Simons Couplings}
\author[a]{Tatsuya Daniel} 
\author[a,b,c]{and Vahid Kamali}

\affiliation[a]{Trottier Space Institute, Department of Physics, McGill University, Montreal, QC H3A 2T8, Canada}
\affiliation[b]{Department of Physics, Bu-Ali Sina (Avicenna) University, Hamedan 65178, 016016, Iran}
\affiliation[c]{School of Continuing Studies, McGill University, Montreal, QC H3A 2T5, Canada}

\emailAdd{tdaniel@physics.mcgill.ca}
\emailAdd{vkamali@ipm.ir}

\abstract{We study the early stages of an oscillatory reheating phase in an inflaton plus spectator axion-SU(2) system, including both an axion-gauge Chern-Simons coupling $\chi F\tilde{F}$ and a gravitational Chern-Simons coupling $\chi R\tilde{R}$. Assuming an isotropic SU(2) background configuration of chromo-natural type and quadratic potentials, we numerically solve the coupled background and tensor perturbation equations during the first e-fold of reheating. The gravitational Chern-Simons term induces a helicity-dependent modification of the tensor kinetic coefficient, yielding a chiral enhancement of the tensor power spectrum on the order of tens of percent for a representative benchmark. We illustrate how such an early-time enhancement can map to a narrow feature in the present-day stochastic gravitational wave spectrum, potentially relevant for upcoming and proposed space-based detectors, while a fully self-consistent determination of the peak scale requires scanning comoving wavenumbers and specifying the reheating history.}


\begin{document}
\maketitle

\section{Introduction}

Axion-gauge field couplings in the early universe have been studied and proposed as a possible source of parity violation, which is observed today in the form of cosmic birefringence, the phenomenon of the linear polarization of cosmic microwave background (CMB) photons being rotated from last scattering to today due to a parity-violating modification to $\Lambda$CDM \cite{Carroll1989,Lepora1998,Minami2020,Komatsu2022,Palazuelos2022}. Such axion-gauge field couplings have been considered in axion inflation, where the axion is coupled to U(1) gauge fields\footnote{The reader is referred to \cite{Pajer2013} for a review on axion inflation.}. The first such theory proposed was called \textit{natural inflation} \cite{Freese1990}; since then, models of axion monodromy inflation have also been popular \cite{Silverstein2008, McAllister2010,Flauger2010,Conlon2012}. The gauge fields can instead be non-Abelian SU(2) fields \cite{Maleknejad2011,Maleknejad2013,Maleknejad2016}, which for example is the case in chromo-natural inflation \cite{Adshead2012,Adshead2013,Adshead2016,Obata2015}.

One way to probe the validity of such couplings in the early universe is by computing the gravitational wave (GW) signal that would be emitted from such a scenario. Thus far, there has been no evidence for a GW background contribution from axion inflation, but it is possible upcoming pulsar timing experiments and space-based GW detectors such as the Square Kilometre Array (SKA) \cite{Dewdney2009} and the Laser Interferometer Space Antenna (LISA) \cite{Amaro_Seoane_2023} may be able to make a detection that could indicate the viability of axion inflation while satisfying other cosmological constraints \cite{Machado2020,Badger2024}. Moreover, the upcoming LiteBIRD \cite{Matsumura2013} and CMB-S4 \cite{CMB-S4:2016} experiments are anticipated to further constrain axion-gauge field couplings.

On the other hand, non-minimal couplings of the axion to the gravitational sector have also been well studied. For example, in a popular extension of general relativity (GR), called Chern-Simons gravity, the axion couples to a Pontryagin term $R\Tilde{R}$, where $R$ is the Ricci curvature tensor  \cite{Jackiw2003,Alexander2009,Yunes2009}. Such a coupling is also parity-violating and has been theorized to give rise to rich phenomena, including gravitational birefringence, where the right- and left-handed modes of GWs propagate differently \cite{Lue1999,Alexander2005,Daniel2024}.

It has been shown that the addition of a gravitational Chern-Simons (GCS) term to an axion-gauge field system during inflation can enhance one helicity of the tensor perturbations by tens of percent. Such a model of inflation has been shown to remain phenomenologically viable \cite{Dimastrogiovanni2017,Mirzagholi2020}, but one can continue to probe such a theory after inflation to test the validity of the model during later cosmological epochs. 

Reheating has been considered for axion-U(1) couplings \cite{Adshead2015,Adshead2024}; \textit{tachyonic preheating} \cite{Felder2001,Kofman2001}, which occurs when the gauge fields are efficiently produced and transfer their energy back to radiation, has been well-studied in this scenario. The dynamics are more complicated when instead considering SU(2) gauge fields. There has been some work on reheating following axion-SU(2) inflation \cite{Fujita2025}, although most work after inflation in this context has typically focused on contributions to dark radiation, as opposed to efficient reheating of the Standard Model (SM) sector \cite{Kakizaki2022}. Thermalization during inflation in axion-SU(2) models has been studied as well \cite{Bhattacharya2025}. However, the extent of previous work involving a model with both GCS and axion-SU(2) couplings is rather limited, especially during reheating; as mentioned earlier, most of the previous work for such theories has been during inflation\footnote{There has been recent broader work on GCS inflation and reheating in a gravitational-anomaly or string-inspired context, rather than focusing exclusively on the chromo-natural SU(2) model, providing alternate UV realizations or motivating different ranges for the GCS coupling \cite{Mavromatos2024,Mavromatos2025,Tzerefos2025}.}.

With this motivation, in this work, we consider the effect of a GCS term to an axion-SU(2) system during the early stages of an oscillatory reheating phase. Due to the observational challenges of realizing such a theory in a minimal chromo-natural-type scenario \cite{Adshead2013,Bagherian2023}, in our theory the axion is a separate spectator field which couples to both the gauge fields and to gravity. We consider an isotropic chromo-natural-type background configuration for the gauge fields, as well as quadratic potentials for both the inflaton and the spectator field at the beginning of reheating. We find that, in the first e-fold of reheating and for a representative benchmark, the coupled tensor evolution results in $\sim 27\%$ enhancement of the left-handed mode and $\sim 14\%$ suppression of the right-handed mode, with a resulting net chirality of about 20\%. This chiral enhancement/suppression produces a GW spectral bump in frequency space, thus possibly being pertinent for future space-based experiments. 

For now, we restrict ourselves to the first e-fold of reheating and do not consider thermalization or an explicit SM sector, no full $k$-scan transfer function (only a single reference mode), and our discussion surrounding space-based detectors is entirely illustrative with order-of-magnitude mappings. We leave a more robust analysis, which will include these topics, for future work.

The outline of this paper is as follows: after laying out the theory in Sec.~\ref{sec:theory}, including the background and perturbations, we detail the setup of numerically solving the system in Sec.~\ref{sec:numerics} before presenting our results in Sec.~\ref{sec:results}. We discuss directions for future work and conclude in Sec.~\ref{sec:discussion}.

Throughout this paper, we use reduced Planck units such that $c = G = \hbar = 1$, unless otherwise noted, and we assume a $(-,+,+,+)$ metric signature; Greek letters ($\mu$,$\nu$,...) range over all spacetime coordinates, Latin letters (i,j,...) range over spatial indices, and derivatives take the form $\partial_\alpha = \frac{\partial}{\partial x^\alpha}$.

\section{Theory}\label{sec:theory}
Our starting point is the action
\begin{align}
    S = S_{\text{EH}} + S_{\varphi} + S_{\text{SPEC}} + S_{\text{GCS}}, \label{eq:starting-action}
\end{align}
where\footnote{We keep the factors of $M_{\text{pl}} = 1/\sqrt{8\pi G}$ explicit for now, setting $M_{\text{pl}}$ to unity beginning in Sec.~\ref{sec:numerics}.} 
\begin{align}
    S_{\text{EH}} &= \int d^4x\sqrt{-g}\frac{M_{\text{pl}}^2}{2}R,
\end{align}
is the usual Einstein-Hilbert term, and the inflaton sector is
\begin{align}
    S_{\varphi} &= \int d^4x\sqrt{-g}\bigg(-\frac{1}{2}(\partial\varphi)^2 - V(\varphi)\bigg),
\end{align}
where $V(\varphi)$ is the inflaton potential. 

\noindent
Moreover, we include the axion $\chi$ as a spectator field, whose action is described by
\begin{align}
    S_{\text{SPEC}} &= \int d^4x\sqrt{-g}\bigg(-\frac{1}{2}(\partial\chi)^2 - U(\chi) - \frac{1}{4}F^a_{\mu\nu}F^{a\mu\nu} + \frac{\lambda_1\chi}{4f}F^a_{\mu\nu}\Tilde{F}^{a\mu\nu}\bigg),
\end{align}
with potential $U(\chi)$, decay constant $f$, coupling $\lambda_1$ to the SU(2) gauge fields, and the gauge field strength tensor
\begin{align}
    F_{\mu\nu}^a = \partial_{\mu}A^a_{\nu} - \partial_{\nu}A^a_{\mu} - g_A\epsilon^{abc}A_{\mu}^bA_{\nu}^c.
\end{align}
Here $g_A$ is the self-coupling constant, $\epsilon^{abc}$ is the three-dimensional Levi-Civita tensor, and $\tilde{F}^{a\mu\nu} \equiv \varepsilon^{\mu\nu\alpha\beta}F^a_{\alpha\beta}/2$ is the dual of $F^a_{\mu\nu}$, where $\varepsilon^{\mu\nu\alpha\beta} \equiv \epsilon^{\mu\nu\alpha\beta}/\sqrt{-g}$ and $\epsilon^{\mu\nu\alpha\beta}$ is the totally anti-symmetric Levi-Civita tensor with $\epsilon^{0123} = 1$.

\noindent 
The axion also couples to the GCS sector\footnote{This is the same theory as in \cite{Mirzagholi2020}, except we will be considering the beginning of reheating as opposed to during inflation.}:
\begin{align}
    S_{\text{GCS}} &= \int d^4x\sqrt{-g}\frac{\lambda_2\chi}{4f}R\Tilde{R},
\end{align}
where  $\lambda_2$ is the coupling of the axion to the Pontryagin term and
\begin{align}
    R\Tilde{R} = R^{\nu~\rho\sigma}_{~\mu}\Tilde{R}^{\mu}_{~\nu\rho\sigma}.
\end{align}

\noindent
$R^{\nu~\rho\sigma}_{~\mu}$ is the Riemann tensor and $\tilde{R}^{\mu}_{~\nu\rho\sigma}$ is its dual, 
\begin{align}
    \Tilde{R}^{\mu}_{~\nu\rho\sigma} = \frac{1}{2}\varepsilon_{\alpha\beta\rho\sigma}R^{\mu~\alpha\beta}_{~\nu}.
\end{align}

\subsection{Background}

In the following discussion for the background and perturbations, we closely follow \cite{Mirzagholi2020}, as the setup is mostly the same except we do not employ any slow-roll approximations corresponding to inflation. 

From Eq.~(\ref{eq:starting-action}), we can find the equations of motion for a Friedmann-Lema\^itre-Robertson-Walker (FLRW) background. The Friedmann equation, as well as the background axion and gauge equations of motion are \cite{Adshead2012, Dimastrogiovanni2017}
\begin{align}
    3M_{\text{pl}}^2\mathcal{H}^2 &= \frac{1}{2}\varphi'^2 + \frac{1}{2}\chi'^2 + a^2V(\varphi) + a^2U(\chi) + \frac{3}{2}\bigg[(Q' + \mathcal{H}Q)^2 + a^2g_A^2Q^4\bigg], \label{eq:friedmann-bkgd} \\
    \chi'' &+ 2\mathcal{H}\chi' + a^2U_{\chi} = -\frac{3g_A\lambda_1}{f}aQ^2(Q' + \mathcal{H}Q), \\
    Q'' &+ 2\mathcal{H}Q' + (\mathcal{H}' + \mathcal{H}^2)Q + 2a^2g_A^2Q^3 = \frac{g_A\lambda_1}{f}a\chi'Q^2, \label{eq:gauge-bkgd}
\end{align}

\noindent
where primes denote derivatives with respect to conformal time $\tau$, $\mathcal{H} \equiv a'/a$ is the conformal Hubble parameter, the subscript $\chi$ denotes a partial derivative with respect to $\chi$, and we have taken an isotropic background gauge field configuration \cite{Maleknejad2011,Maleknejad2013},
\begin{align}
    A_0^a = 0,~~A_i^a = \delta^a_ia(\tau)Q(\tau).
\end{align}

Since $R\tilde{R}$ vanishes in an FLRW background, the GCS term does not contribute to the background equations of motion.

\subsection{Perturbations}

Upon perturbing the FLRW metric as
\begin{align}
    ds^2 = a^2(\tau)[-d\tau^2 + (\delta_{ij} + \Tilde{h}_{ij})dy^idy^j],
\end{align}
we can define the Fourier transformed right- and left-handed circular polarization states as
\begin{align}
    \Tilde{h}_{ij}(\tau,y) = \int \sum_{R,L}\frac{d^3k}{(2\pi)^{3/2}}e_{ij}^A(k)\Tilde{h}_A(\tau,k)e^{iky},
\end{align}
where $e_{ij}^A$ is the polarization tensor for the right- and left-handed circular polarization states (denoted $A = R,L$, respectively) and satisfies
\begin{align}
    ik_a\epsilon^{ab}_{~~c}e^R_{db} = ke^R_{cd},~~ik_a\epsilon^{ab}_{~~c}e^L_{db} = -ke_{cd}^L,
\end{align}
with $\epsilon^{ab}_{~~c}$ again being the three-dimensional Levi-Civita symbol. We will for simplicity assume that the GWs are propagating in the $z$-direction,
\begin{align}
    ds^2 = a^2(\tau)\bigg[-d\tau^2 + (1 + \Tilde{h}_+(\tau,z))dx^2 + (1 - \Tilde{h}_+(\tau,z))dy^2 + 2\Tilde{h}_\times(\tau,z)dxdy + dz^2\bigg].
\end{align}

We will write the tensor perturbations of the gauge field as $\delta A_i^a = a\Tilde{t}^a_i$, where we choose $\Tilde{t}^a_i$ to be transverse and traceless ($\partial_i\Tilde{t}^a_i = \Tilde{t}^{ai}_i$ = 0). Moreover, we will write the gauge tensor perturbations as $\delta A_i^1 = a(\Tilde{t}_+,\Tilde{t}_{\times},0)$ and $\delta A_i^2 = a(\Tilde{t}_{\times},-\Tilde{t}_+,0)$, and we define the canonically normalized tensor perturbations
\begin{align}
    h_{ij} \equiv a\frac{M_{\text{pl}}}{\sqrt{2}}\Tilde{h}_{ij},~~t^a_i \equiv \sqrt{2}a\Tilde{t}^a_i. \label{eq:h-t-normalization}
\end{align}

\noindent
The right- and left-handed helicities are then defined as
\begin{align}
    h_{R,L} \equiv \frac{1}{\sqrt{2}}(h_+ \mp ih_{\times}),~~t_{R,L} \equiv \frac{1}{\sqrt{2}}(t_+ \mp it_{\times}). \label{eq:h-and-t}
\end{align}

\noindent
Writing the action Eq.~(\ref{eq:starting-action}) to second order in the perturbations, we have \cite{Mirzagholi2020}
\begin{align}
    S &= \frac{1}{2}\sum_{A = R,L}\int d\tau d^3k\bigg\{\bigg(1 - \frac{s\lambda_2\chi'}{4faM_{\text{pl}}^2}\frac{k}{a}\bigg)\bigg[h_A'^{\dagger}h_A' + (-k^2 + 2\mathcal{H}^2)h_A^{\dagger}h_A - 2\mathcal{H}\mathfrak{R}(h_A'^{\dagger}h_A)\bigg] \nonumber \\ &+ t_A^{\dagger} t_A\bigg[sak\bigg(2g_AQ + \frac{\lambda_1}{f}a\chi'\bigg) - k^2 - \frac{g_AaQ\lambda_1}{f}\chi'\bigg] + 2\mathfrak{R}\bigg[h_A'^{\dagger}t_A - h_A^{\dagger}t_A'\bigg]\bigg(\frac{Q' + \mathcal{H}Q}{M_{\text{pl}}}\bigg) \nonumber \\ &+ t_A^{\dagger}t_A' + 2\mathfrak{R}(h_At_A^{\dagger})\bigg[-sak\frac{2g_AQ^2}{M_{\text{pl}}} - \frac{\mathcal{H}(Q' + \mathcal{H}Q)}{M_{\text{pl}}} + \frac{g_A\lambda_1Q^2a}{fM_{\text{pl}}}\chi'\bigg], \label{eq:action}
\end{align}
where $s = \pm 1$ for $A = R,L$ respectively, and again primes denote derivatives with respect to conformal time.

\noindent 
From Eq.~(\ref{eq:action}), we can write the action density as
\begin{align}
    \mathcal{L} &= \mathcal{A}(k,\tau)\bigg[h_A'^{\dagger}h_A' + (-k^2 + 2\mathcal{H}^2)h_A^{\dagger}h_A - 2\mathcal{H}\mathfrak{R}(h_A'^{\dagger}h_A)\bigg] + t_A'^{\dagger}t_A' + t_A^{\dagger}t_A\mathcal{S}(k,\tau) \nonumber \\ &+ 2\mathfrak{R}[h_A'^{\dagger}t_A - h_A^{\dagger}t_A']B(\tau) + 2\mathfrak{R}(h_At_A^{\dagger})\mathcal{C}(k,\tau), \label{eq:lagrangian}
\end{align}
where
\begin{align}
    \mathcal{A}(k,\tau) &= 1 - s\frac{\lambda_2}{4faM_{\text{pl}}^2}\frac{k}{a}\chi', \label{eq:A} \\
    B(\tau) &= \frac{Q' + \mathcal{H}Q}{M_{\text{pl}}}, \\
    \mathcal{S}(k,\tau) &= sak\bigg(2g_AQ + \frac{\lambda_1}{f}a\chi'\bigg) - k^2 - g_AaQ\frac{\lambda_1}{f}\chi', \\
    \mathcal{C}(k,\tau) &= \frac{g_A\lambda_1}{fM_{\text{pl}}}aQ^2\chi' - \frac{2sakg_AQ^2}{M_{\text{pl}}} - \frac{\mathcal{H}(Q' + \mathcal{H}Q)}{M_{\text{pl}}}. \label{eq:C}
\end{align}

\noindent
The tensor equations of motion are then
\begin{align}
    \mathcal{A}(x)\partial_x^2h_A &- \frac{\partial_x\mathcal{A}}{k}\partial_xh_A - \frac{B(x)}{k}\partial_xt_A + \frac{\partial_xB}{k^2}t_A \nonumber \\ &- \frac{1}{k^2}\bigg\{\mathcal{A}(x)\bigg[-k^2 + 2\mathcal{H}^2(x)\bigg] + (\partial_x\mathcal{A})\mathcal{H}(x) + \mathcal{A}(x)\partial_x\mathcal{H}\bigg\}h_A = 0, \label{eq:hA-tensor-eom} \\
    \partial_x^2t_A &- \frac{\mathcal{S}(x)}{k^2}t_A + \frac{2B(x)}{k}\partial_xh_A - \frac{\partial_xB}{k^2}h_A - \frac{\mathcal{C}(x)}{k^2}h_A = 0, \label{eq:tA-tensor-eom}
\end{align}

\noindent
where we have defined $x \equiv -k\tau$, and the subscript $x$ denotes a partial derivative with respect to $x$.

We observe that $\mathcal{A}(k,\tau)$, as defined in Eq.~(\ref{eq:A}), is helicity-dependent and largest when $\dot{\chi} = \chi'/a$ and $k_{\text{phys}} = k/a$ are large. As such, the GCS effect will be early-time dominated, but eventually $k_{\text{phys}}$ will cause a violation of the ghost-free bound. Specifically, we note that the theory is healthy only if the kinetic coefficient ${\cal A}(x)$ is positive for each helicity:
\begin{align}
{\cal A}_L(x)>0,\qquad {\cal A}_R(x)>0.
\end{align}

In Appendix~\ref{sec:ghost-free}, we show that ${\cal A}(x)$ remains positive for both helicities (so no ghost), while deviating from unity enough to affect the tensor evolution.

Eqs.~(\ref{eq:hA-tensor-eom}) and (\ref{eq:tA-tensor-eom}), combined with Eqs.~(\ref{eq:friedmann-bkgd})-(\ref{eq:gauge-bkgd}), make up the system of equations that we will solve.

\section{Numerics}\label{sec:numerics}
In this section, we detail our numerical setup for solving the system of background and perturbation equations in Sec.~\ref{sec:theory}.

\subsection{Parameters and Initial Conditions}\label{sec:params-and-ics}
Setting $M_{\text{pl}} = 1$, and again working with the dimensionless variable $x \equiv -k\tau$, we set the reference scale to $k = 1$, and we take the reheating potentials to be quadratic,
\begin{align}
    V(\varphi) = \frac{1}{2}m^2\varphi^2,~~U(\chi) = \frac{1}{2}m^2\chi^2.
\end{align}
Moreover, we introduce the auxiliary variables \cite{Mukhanov2005,Weinberg2008}
\begin{align}
\pi_{\varphi} \equiv~&\varphi',~~~\pi_{\chi} \equiv \chi',~~~\pi_Q \equiv Q', \nonumber \\
\Pi_\varphi(x)\equiv a^2\varphi_x,\qquad
&\Pi_\chi(x)\equiv a^2\chi_x,\qquad
{\cal Q}(x)\equiv a\,Q,\qquad
\Pi_{\cal Q}(x)\equiv \partial_x{\cal Q}. \label{eq:aux-variables}
\end{align}
Our parameter choices and initial conditions are shown in Table~\ref{tab:params}. 
We solve the coupled background system, Eqs.~(\ref{eq:friedmann-bkgd})-(\ref{eq:gauge-bkgd}), together with the tensor perturbation equations, Eqs.~(\ref{eq:hA-tensor-eom}) and (\ref{eq:tA-tensor-eom}), by rewriting them as a first-order system (see Appendix~\ref{sec:first-order-diffeqs}) and integrating from $x_{\rm reh}$ to $x_{\rm end}$. Shifting $\tau\to\tau-\tau_\star$
so that $x>0$ during the reheating interval of interest\footnote{Since $x\equiv -k\tau$ is defined up to a constant shift $\tau\to\tau-\tau_\star$ (and $d/d\tau=-k\,d/dx$),
a constant shift of $x$ does not change the equations of motion. }, we use a Dormand-Prince 8(5,3) adaptive Runge-Kutta integrator \cite{Dormand1980}, with internal step sizes chosen automatically to satisfy the requested tolerances. 

We verify numerical convergence by (i) tightening tolerances by an order of magnitude, (ii) refining the output grid, and (iii) repeating the tensor evolution using the \texttt{RK45} \cite{Hairer1993} and \texttt{RK23} \cite{Bogacki1989} methods, finding that these adjustments change the final tensor power ratios by less than $\mathcal{O}(10^{-6})$. We additionally check that the oscillatory system does not exhibit stiffness, and that shifting the initial time deeper inside the horizon leaves the final power ratios unchanged up to numerical roundoff. More details on the numerical integration can be found in Appendix~\ref{sec:numerical-integration}.

\begin{table}[t]
\centering
\begin{tabular}{c|c}
\hline
Parameter & Value \\
\hline
$M_{\rm pl}$ & $1$ \\
$k$ & $1$ \\
$f$ & $1$ \\
$m_\varphi$ & $0.8$ \\
$m_\chi$ & $0.5$ \\
$g_A$ & $0.2$ \\
$\lambda_1$ & $20$ \\
$\lambda_2$ (benchmarks) & $0,\ 100$ \\
\hline
Initial $x_{\rm reh}$ & $50$ \\
Initial $a(x_{\rm reh})$ & $1$ \\
Initial $\varphi(x_{\rm reh})$ & $0.08$ \\
Initial $\pi_{\varphi}(x_{\rm reh})$ & $0$ \\
Initial $\Pi_\varphi(x_{\rm reh})$ & $0$ \\
Initial $\chi(x_{\rm reh})$ & $-0.05$ \\
Initial $\pi_\chi(x_{\rm reh})$ & $0$ \\
Initial $\Pi_\chi(x_{\rm reh})$ & $0$ \\
Initial ${\cal Q}(x_{\rm reh})$ & $0.05$ \\
Initial $\pi_Q(x_{\rm reh})$ & $-0.05\mathcal{H}_{\text{reh}}$ \\
Initial $\Pi_{\cal Q}(x_{\rm reh})$ & $0$ \\
\hline
\end{tabular}
\caption{Our choice of parameters and initial conditions. The parameters $m_{\varphi},m_{\chi},g_A$, $\lambda_1$ and $\lambda_2$ are chosen to illustrate representative weak-coupling regimes typical of axion-gauge inflationary models (e.g. \cite{Adshead2012,Maleknejad2013}). The initial conditions are specified with small field amplitudes and vanishing canonical momenta (except for $\pi_Q$; $\mathcal{H}_{\text{reh}}$ is determined from the Friedmann equation (\ref{eq:friedmann-bkgd}) using the initial field values), while the initial time $x_{\text{reh}} \gg 1$ ensures that the perturbation modes begin well within the horizon. 
The spectator sector is chosen to be energetically subdominant at $x_{\rm reh}$.
}
\label{tab:params}
\end{table}

\subsection{Mode Functions}
Because our tensor system is coupled, the vacuum of the quantized solution contains two independent mode-function sectors, which we will call the ``metric vacuum'' and ``gauge vacuum,'' respectively. To isolate the contributions from the metric and gauge fluctuations, we numerically evolve two sets of initial conditions at $x_{\rm reh}$, corresponding to modes initialized well within the horizon in the Bunch-Davies vacuum \cite{Weinberg2008,Dodelson2003}:
\begin{align}
\text{(i) metric vacuum:}\quad & h_A(x_{\rm reh})=\frac{e^{ix_{\rm reh}}}{\sqrt{2k}},\quad
\partial_xh_A(x_{\rm reh})=i\,h_A(x_{\rm reh}),\quad
t_A=\partial_xt_A=0, \label{eq:metric-vacuum-mode-ic}
\\
\text{(ii) gauge vacuum:}\quad & t_A(x_{\rm reh})=\frac{e^{ix_{\rm reh}}}{\sqrt{2k}},\quad
\partial_xt_A(x_{\rm reh})=i\,t_A(x_{\rm reh}),\quad
h_A=\partial_xh_A=0. \label{eq:gauge-vacuum-mode-ic}
\end{align}

At sufficiently large $x$, the mixing terms are subdominant to $k^2$, so the eigenmodes approach free Bunch-Davies modes. We then obtain the physical metric tensor power\footnote{As we are mainly interested in studying the chiral contributions to the GW power, here our final tensor mode ratios will only include the metric power from $h_A$ and not the gauge power from $t_A$, i.e. we do not add $|t_A|^2$ to define ``total power." However, because the tensor system is coupled, the metric perturbation will have contributions from both the usual vacuum perturbation and the gauge-sourced piece.} by summing the contributions from both sectors\footnote{Eq.~(\ref{eq:metric-tensor-power}) follows from using that $P_h^A = |\tilde{h}_{A,{\rm (i)}}|^2+|\tilde{h}_{A,{\rm (ii)}}|^2$, in our convention, and using Eq.~(\ref{eq:h-t-normalization}) to rewrite $P_h^A$ in terms of $h_A$. The dimensionless power spectrum $\mathcal{P}_h^A$ is then defined as $\mathcal{P}_h^A \equiv (k^3/2\pi^2)P_h^A$. As we will just be comparing ratios, all prefactors will cancel.},
\begin{align}
    P_h^A = \frac{2}{a^2}\bigg(|h_{A,{\rm (i)}}|^2+|h_{A,{\rm (ii)}}|^2\bigg). \label{eq:metric-tensor-power}
\end{align}

\section{Results}\label{sec:results}

\subsection{Scalar field oscillations and equation of state}
\begin{figure}[t]
\centering
\includegraphics[width=\linewidth]{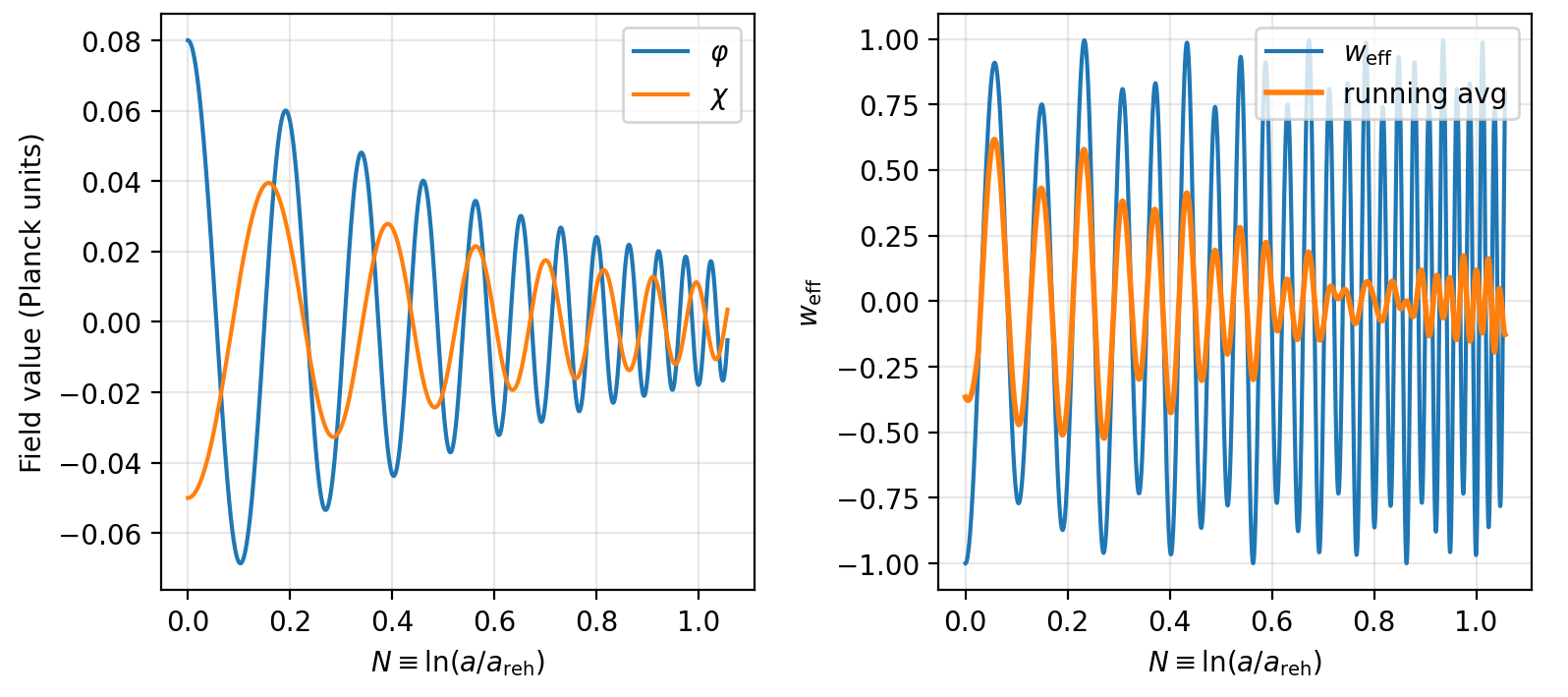}
\caption{Background reheating dynamics for the parameter choice and initial conditions in Table~\ref{tab:params}.
\textbf{Left:} oscillations of the inflaton $\varphi$ and spectator axion $\chi$. \textbf{Right:} effective equation of state $w_{\rm eff}=p_{\rm tot}/\rho_{\rm tot}$ and the short running average; we see that$\langle w\rangle\simeq 0$.
}
\label{fig:bkgd-evolution}
\end{figure}

\begin{figure}[t]
\centering
\includegraphics[width=0.5\linewidth]{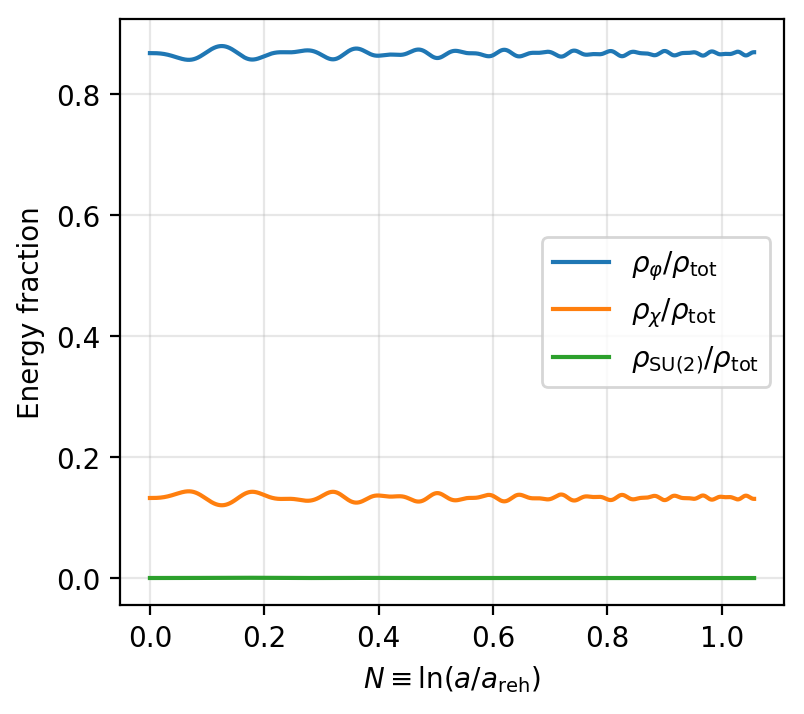}
\caption{Energy fractions showing that the spectator axion--SU(2) sector remains subdominant, for the parameter choice in Table~\ref{tab:params}.
}
\label{fig:bkgd-evolution-2}
\end{figure}

With quadratic potentials, a homogeneous scalar field oscillating about the minimum of its potential has an average equation of state $\langle w\rangle\simeq 0$,
so the expansion is effectively matter-dominated, up to oscillations and subdominant sectors.
Figs.~\ref{fig:bkgd-evolution} and \ref{fig:bkgd-evolution-2} show the background evolution: $\varphi$ and $\chi$ oscillate, the effective $w_{\rm eff}$ oscillates about zero,
and the inflaton energy density remains dominant.

\begin{figure}[t]
\centering
\includegraphics[width=\linewidth]{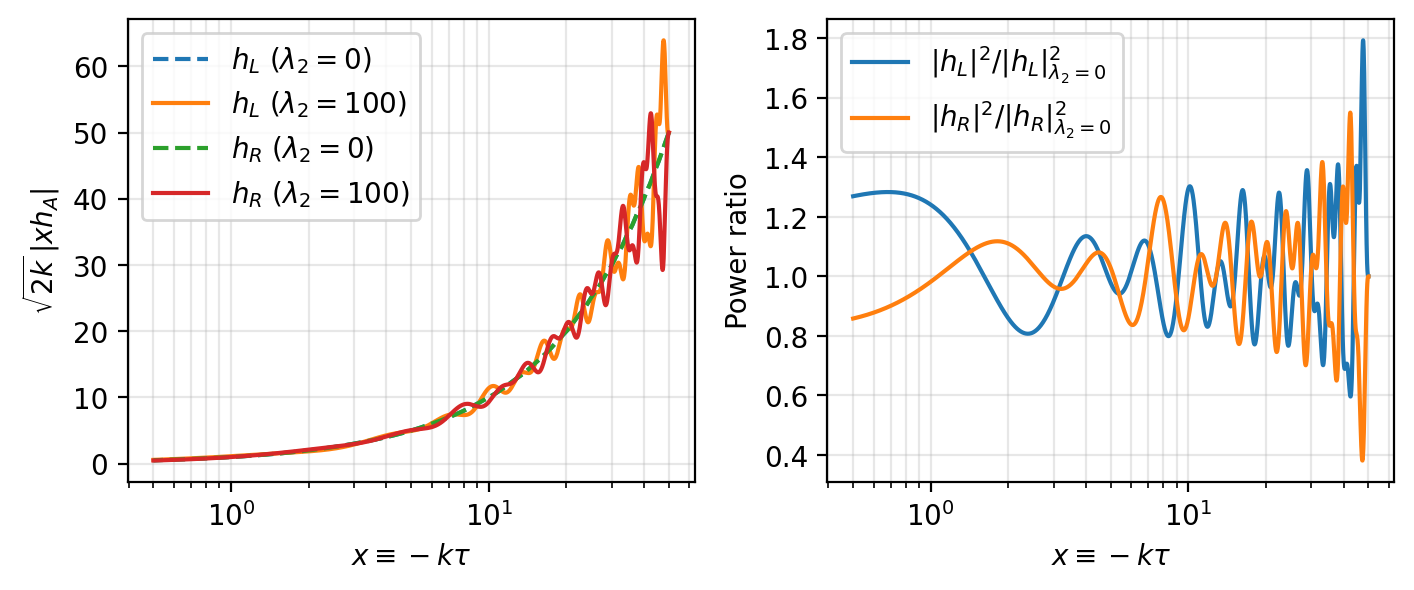}
\caption{Tensor mode evolution during reheating, using the parameter values in Table~\ref{tab:params}. \textbf{Left:} $\sqrt{2k}\,|x\,h_A|$ for $A=R,L$ with (solid) and without (dashed) the GCS coupling. \textbf{Right:} power ratios relative to $\lambda_2=0$.
The GCS interaction produces a chiral enhancement that accumulates over the reheating evolution.}
\label{fig:tensors}
\end{figure}

\subsection{Chiral tensor enhancement}\label{sec:chiral-tensor-enhancement}
Fig.~\ref{fig:tensors} shows the tensor mode evolution for $\lambda_2=0$ and $\lambda_2=100$.
We see that the GCS coupling amplifies the left-handed tensor mode $h_L$ during reheating and suppresses the right-handed mode $h_R$. At the end of the numerical reheating interval ($x_{\rm end}=0.5$)\footnote{$x_{\text{end}}$ is chosen such that the integration lasts for sufficiently long (one e-fold) to obtain a noticeable GCS enhancement, but not before we run into the instability bound. If we were to run the integration for more than one e-fold, we expect the GCS enhancement/suppression ratios in Eq.~(\ref{eq:tensor-power-ratios}) to change, but (i) we would soon run into an instability ($\mathcal{A} < 0$), and (ii) one would need to consider the matter and radiation components that have been produced, thus resulting in different background/perturbation equations (more discussed in Sec.~\ref{sec:discussion}).\label{footnote:x-end-justification}}, for $h_A$ we find
\begin{align}
\frac{P_h^L(\lambda_2=100)}{P_h^L(\lambda_2=0)}\simeq 1.27,\qquad
\frac{P_h^R(\lambda_2=100)}{P_h^R(\lambda_2=0)}\simeq 0.86, \label{eq:tensor-power-ratios}
\end{align}
which corresponds to a net chirality
\begin{align}
\Delta_\chi\equiv \frac{P_h^R-P_h^L}{P_h^R+P_h^L}\simeq -0.19,
\end{align}
i.e. a left-handed excess generated during reheating.

\subsection{Stochastic GW background}
For modes that re-enter the horizon during radiation domination, a standard approximation relates the present-day GW energy density to the
primordial tensor power spectrum $P_t(k)$ via \cite{Boyle2008,Caprini2018}
\begin{align}
\Omega_{\rm GW,0}(k)\simeq \frac{\Omega_{r,0}}{24}\,P_t(k)\,,
\label{eq:OmegaGW_simple}
\end{align}
where $\Omega_{r,0}$ is the present-day radiation fraction\footnote{Here we are not considering the detailed transfer function from the matter era or changes in relativistic degrees of freedom; this is sufficient for an order-of-magnitude detectability estimate in the frequency bands we are interested in.}.
The comoving wavenumber is related to the physical frequency today by \cite{Maggiore2000,Bartolo2016}
\begin{align}
f = \frac{k}{2\pi a_0}\simeq 1.5\times 10^{-15}\,{\rm Hz}\left(\frac{k}{1~{\rm Mpc}^{-1}}\right). \label{eq:f-k-rel}
\end{align}

Because the GCS enhancement in our setup is generated during the first e-fold of reheating\footnote{We are not claiming that the GCS enhancement will be generated \emph{only} during the first e-fold of reheating. We have not yet considered what happens after the first e-fold of reheating, for the reasons mentioned in footnote~\ref{footnote:x-end-justification} and discussed in Sec.~\ref{sec:discussion}.}, the resulting feature in $P_t(k)$ is naturally localized in $\ln k$.
A useful phenomenological parameterization is therefore \cite{Caprini2018,Domenech2021}
\begin{align}
P_t(k)=A_s\left[r_{\rm vac} + r_\star\exp\!\left(-\frac{\ln^2(k/k_\star)}{2\sigma^2}\right)\right],
\label{eq:Pt_param}
\end{align}
where $A_s$ is the scalar amplitude at CMB scales, $r_{\rm vac}$ is the usual (nearly) scale-invariant vacuum contribution, and the second term models a reheating-generated bump centered at $k_\star$ with width $\sigma$ and enhancement amplitude $r_\star$.

\noindent
In the figures below, we take the following \cite{Planck2018,Ade2021}\footnote{We adopt a value for $r_{\text{vac}}$ that is below the current CMB bound of $r < 0.036$ \cite{Ade2021}. The enhancement amplitude $r_\star = 200$ is purely phenomenological, representing a strongly-sourced GW contribution. For a matter-like reheating phase ($w\simeq 0$), one e-fold in scale factor corresponds to $\Delta\ln k\simeq -1/2$, which motivates taking $\sigma={\cal O}(0.5)$.}:
\begin{align}
A_s = 2.1\times 10^{-9},\quad \Omega_{r,0}=9.2\times 10^{-5},\quad r_{\rm vac}=4.6\times 10^{-3},\quad r_\star=200,\quad \sigma=0.5.
\end{align}

To obtain a self-consistent value for the peak frequency $f_\star$, which determines where the reheating-generated feature sits in frequency space, one would need to specify the energy scale at the end of inflation and reheating history (e.g. equation of state and duration), evolve the coupled tensors for a range of comoving modes $k$, and then identify the wavenumber $k_\star$ at which the reheating-era GCS enhancement peaks. $f_\star$ would then correspond to redshifting $k_\star$ to today, via Eq.~(\ref{eq:f-k-rel}). 

We do not attempt a first-principles prediction here for the peak scale $k_\star$ of the reheating-generated feature. We simply seek to use the phenomenological parameterization Eq.~(\ref{eq:Pt_param}) to demonstrate how the reheating-era chiral amplification computed in Sec.~\ref{sec:chiral-tensor-enhancement} acts as an approximately multiplicative, helicity-dependent rescaling of the amplitude. Thus, for now we will take $f_\star$ as an input parameter, leaving a self-consistent determination of $f_\star$ for future work.

\begin{figure*}[htb!]
    \includegraphics[width=0.5\textwidth]{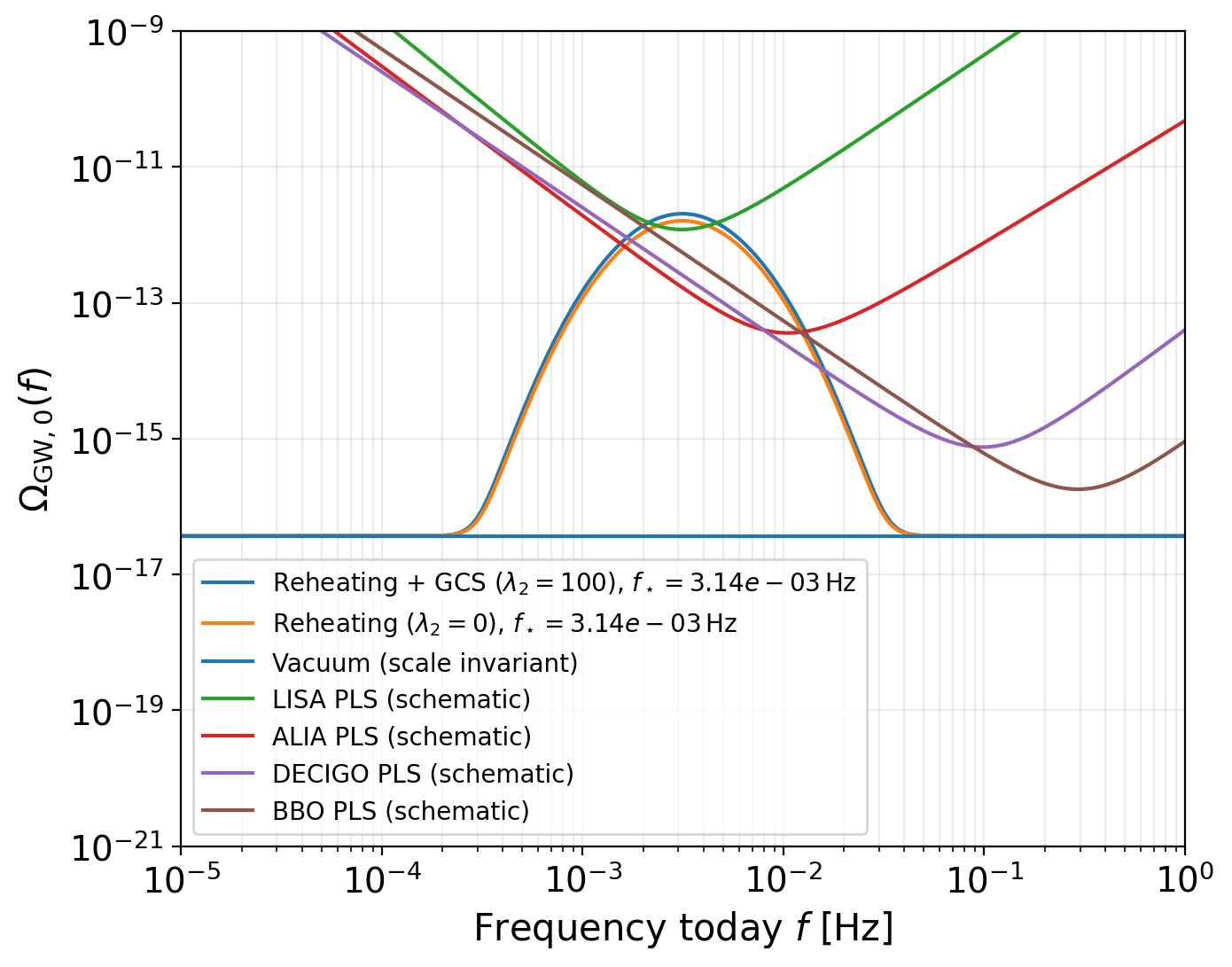}\hfill
    \includegraphics[width=0.5\textwidth]{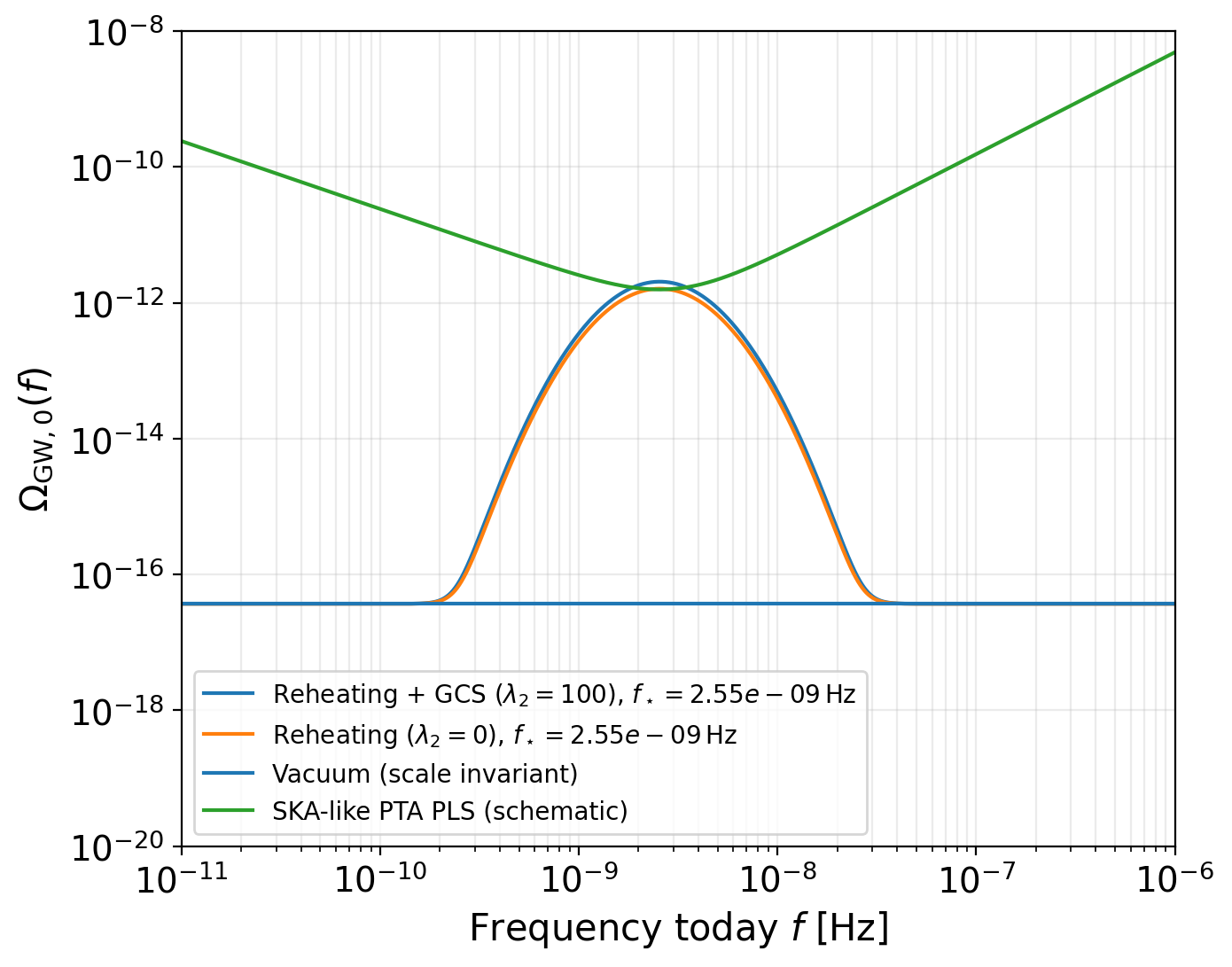}
    \caption{The present-day $\Omega_{\rm GW,0}(f)$ from Eqs.~\eqref{eq:OmegaGW_simple} and \eqref{eq:Pt_param}, using the parameters and initial conditions in Table~\ref{tab:params}.
Unlike inflationary production, the enhancement is localized because it is generated during the first e-fold of reheating. \textbf{Left:} Illustrative sensitivities of LISA, the Advanced Laser Interferometer Antenna (ALIA), the Deci-hertz Interferometer Gravitational Wave Observatory (DECIGO) and Big Bang Observatory (BBO) in an order-of-magnitude detectability sketch.
The LISA curve shows an approximate power-law integrated sensitivity (PLS) curve (SNR=10, 4-year mission), using the noise model of \cite{Robson2019} and the PLS construction in \cite{Caprini2019}. Sensitivity curves for ALIA, DECIGO and BBO follow the mission concepts described in \cite{Crowder2005,Seto2001,Kawamura2011,Corbin2006}. The bump amplitude/width and peak frequency are chosen for demonstration rather than derived from a full $k$-scan of the reheating dynamics. The plotted effect should therefore be interpreted as an order-unity rescaling of the bump amplitude, via Eq.~(\ref{eq:tensor-power-ratios}), rather than as a parameter-free forecast. 
\textbf{Right:}
Same as the left panel, but for a SKA-like PTA, using a simplified white-noise model and the analytic sensitivity formalism of \cite{Moore2015}.} 
    \label{fig:sgwb-closeup}
\end{figure*}

\begin{figure}[t]
\centering
\includegraphics[width=0.7\linewidth]{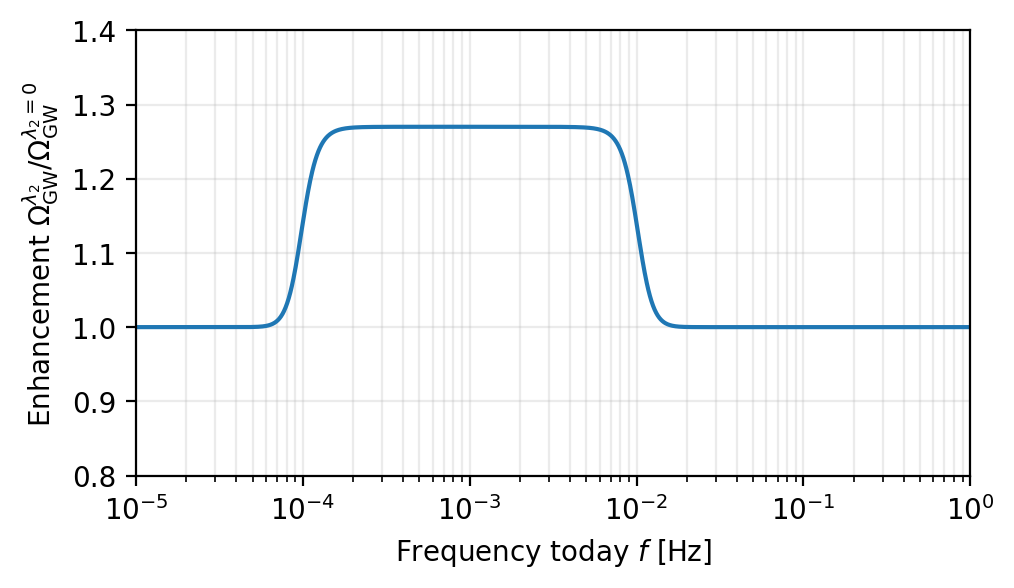}
\caption{Reheating-era GCS enhancement as a frequency-dependent factor for the LISA-band benchmark.
For a narrow reheating bump, the GCS term effectively rescales the peak amplitude of the dominant helicity by a factor of order unity (here $\sim1.27$), leaving the spectrum unchanged away from the reheating feature.
}
\label{fig:enhFactor}
\end{figure}

We show two choices of the peak frequency: $f_\star= 3.14 \times 10^{-3}\,{\rm Hz}$, to target the LISA band, and $f_\star=2.55\times 10^{-9}\,{\rm Hz}$, to target the pulsar timing array (PTA)/SKA band. These correspond roughly to the frequencies at which LISA and SKA are most sensitive.
As a rough mapping between a characteristic early-universe scale and the observed frequency, modes entering the horizon during radiation domination at temperature $T_\star$ correspond to \cite{Maggiore2000,Watanabe2006}
\begin{align}
f \simeq 1.3\times 10^{-7}\,{\rm Hz}\left(\frac{T_\star}{1~{\rm GeV}}\right)\left(\frac{g_\star}{100}\right)^{1/6},
\end{align}
so $f_\star\simeq 10^{-3}\,{\rm Hz}$ corresponds to $T_\star={\cal O}(10~{\rm TeV})$.

We plot $\Omega_{\rm GW,0}(f)$ obtained by combining Eqs.~\eqref{eq:OmegaGW_simple} and \eqref{eq:Pt_param} in Fig. \ref{fig:sgwb-closeup}. Unlike inflationary spectator scenarios, which produce an extended enhancement over many decades \cite{Mirzagholi2020,Komatsu2022}, reheating produces a narrow frequency bump. To emphasize the size of the parity-violating correction, Fig.~\ref{fig:enhFactor} plots the enhancement factor for the illustrative LISA-band benchmark.
The peak enhancement tracks the $\simeq 1.27$ tensor-power ratio extracted from Fig.~\ref{fig:tensors}.

\section{Discussion and Conclusion}\label{sec:discussion}
In this paper, we have investigated an axion-SU(2) system with a gravitational Chern-Simons (GCS) coupling during the early stages of reheating, where the axion, serving as a spectator field, couples to both the gauge fields and the Pontryagin term $R\Tilde{R}$ that appears in Chern-Simons gravity. For a representative benchmark, we have found that the GCS term enhances the left-handed mode by $\sim 27\%$ and suppresses the right-handed mode by $\sim 14\%$, with a resulting net chirality of about $20\%$. This yields a narrow GW spectrum in frequency space, which is potentially relevant for upcoming and proposed space-based detectors. A more complete analysis is left for future work, some points for which we discuss below.

As we are only considering the first e-fold of reheating, and we are not assuming the reheating to occur instantaneously, we do not consider any other radiation or matter components in the universe. At some point after the first e-fold of reheating, the produced radiation and matter will no longer be negligible, and one would need to use cosmological perturbation theory to derive the full set of background and perturbation equations governing such a system in order to properly determine its full evolution. This analysis could be useful if one wishes to, for example, consider the fate of the axion at later times from this model, possibly as a candidate for dark matter or dark energy, or consider additional contributions to the stochastic GW background that could be produced from this model after reheating. We leave this for future work. 

Furthermore, the GCS coupling mainly affects GWs, not the energy transfer to the Standard Model (SM) plasma directly unless the axion is coupled to SM fields. Because GCS terms don't naturally produce SM particles during reheating, the SU(2) gauge sector would need additional couplings to the SM for thermalization, such as to fermions or a Higgs portal. Whether and how the spectator axion can efficiently transfer its energy to the SU(2) or SM sectors after inflation, in the presence of GCS couplings, remains an open question.

Most of the contribution of the GCS coupling to the tensor enhancement occurs at the beginning of reheating. This is because the GCS term affects the kinetic coefficient $\mathcal{A}$ of the tensor modes, so its effect is largest when the axion velocity is sizable and the physical wavenumber is large enough to produce a non-negligible addition, but not so large that $\mathcal{A}$ becomes negative and we introduce a ghost. In our reheating setup, these conditions are naturally met early, before the axion oscillations redshift away and before the physical wavenumber grows to the point that the ghost bound becomes too restrictive.
This is why a first-e-fold window is a sensible place to search for the effect. When analyzing this theory in the later stages of reheating, one would need to consider shutting off the GCS coupling before it introduces a ghost.

We close by noting other directions for future work include (i) mapping the end of inflation in chromo-natural-type models to the initial conditions at the beginning of reheating, thus assessing the sensitivity of these results to the initial conditions, (ii) scanning a range of comoving $k$ in order to obtain a full transfer function from reheating for the tensor spectrum, as well as (iii) considering inflaton decay during the later stages of reheating, and checking backreaction from the produced gauge or tensor perturbations.

\acknowledgments
The authors thank Robert Brandenberger and Yong Xu for useful discussions, including helpful comments on a draft of this work. T.D. is supported by a Trottier Space Institute fellowship and by funds from NSERC. V.K. would like to acknowledge the McGill University Physics Department for hospitality.

\appendix

\section{First-Order Differential Equations} \label{sec:first-order-diffeqs}
\subsection{Background}
Using the auxiliary variables that we have defined in Eq.~(\ref{eq:aux-variables}), the Friedmann equation takes the form 
\begin{align}
    3M_{\text{pl}}^2\mathcal{H}^2 = \frac{1}{2}\pi_{\varphi}^2 + \frac{1}{2}\pi_{\chi}^2 + a^2V(\varphi) + a^2U(\chi) + \frac{3}{2}\bigg[(\pi_Q + \mathcal{H}Q)^2 + a^2g_A^2Q^4\bigg]. \label{eq:bg_fr}
\end{align}

\noindent
The background axion and gauge field equations can be written as
\begin{align}
    \pi_{\chi}' &= -2\mathcal{H}\pi_{\chi} - a^2U_{\chi} - \frac{3g_A\lambda_1}{f}aQ^2(\pi_Q + \mathcal{H}Q), \label{eq:bg_chi_firstorder} \\
    \pi_Q' &= -2\mathcal{H}\pi_Q - (\mathcal{H}' + \mathcal{H}^2)Q - 2a^2g_A^2Q^3 + \frac{g_A\lambda_1}{f}a\pi_{\chi}Q^2, \label{eq:bg_Q_firstorder}
\end{align}
with the kinematic equations
\begin{align}
    a' = \mathcal{H}a,~~~\varphi' = \pi_{\varphi},~~~\chi' = \pi_{\chi}',~~~Q' = \pi_Q.
\end{align}

\subsection{Perturbations}
After rewriting the second-order perturbation system as a first-order system by defining
\begin{align}
u_A(x)\equiv \partial_x h_A(x),\qquad v_A(x)\equiv \partial_x t_A(x),
\end{align}
the tensor equations (\ref{eq:hA-tensor-eom}) and (\ref{eq:tA-tensor-eom}) become
\begin{align}
\partial_x h_A &= u_A,
\nonumber\\
\partial_x u_A &=
\frac{\partial_x{\cal A}}{k{\cal A}}\,u_A
+\frac{B}{k{\cal A}}\,v_A
-\frac{\partial_x B}{k^2{\cal A}}\,t_A
+\frac{1}{k^2{\cal A}}
\bigg[
{\cal A}\big(-k^2+2{\cal H}^2\big)
+(\partial_x{\cal A}){\cal H}
+{\cal A}\,\partial_x{\cal H}
\bigg]h_A,
\label{eq:tensor_firstorder_h}
\\
\partial_x t_A &= v_A,
\nonumber\\
\partial_x v_A &=
\frac{{\cal S}}{k^2}\,t_A
-\frac{2B}{k}\,u_A
+\frac{1}{k^2}\bigg(\partial_x B+ \mathcal{C}\bigg)h_A,
\label{eq:tensor_firstorder_t}
\end{align}
where $\mathcal{A}, B, \mathcal{C}$ and $\mathcal{S}$ are given by Eqs.~(\ref{eq:A})-(\ref{eq:C}).

\section{Numerical Integration} \label{sec:numerical-integration}
We solve both the reheating background system and the coupled tensor system
using the adaptive, high-order explicit Runge--Kutta integrator
\texttt{DOP853} (Dormand--Prince 8(5,3)) \cite{Dormand1980}.
We do not impose a fixed step size; step sizes are chosen adaptively to meet the requested error controls.
We choose the following:
\begin{align}
  \text{background:}\quad & \texttt{rtol}=10^{-9},\ \texttt{atol}=10^{-12},\\
  \text{perturbations:}\quad & \texttt{rtol}=10^{-8},\ \texttt{atol}=10^{-10}.
\end{align}
Convergence was checked by tightening tolerances by an order of magnitude, as well as refining the output grid (\texttt{n\_eval} = 6000 $\to$ 8000 for the background, \texttt{n\_eval} = 3000 $\to$ 4000 for the perturbations) and repeating the tensor evolution with the \texttt{RK45} \cite{Hairer1993} and \texttt{RK23} \cite{Bogacki1989} methods. We confirmed that
the late-time enhancement/suppression ratios $P_h^{A}(\lambda_2=100)/P_h^{A}(\lambda_2=0)$ in Eq.~(\ref{eq:tensor-power-ratios}) change by $\mathcal{O}(10^{-7}-10^{-9})$ with these adjustments.

\begin{figure}[t]
\centering
\includegraphics[width=\linewidth]{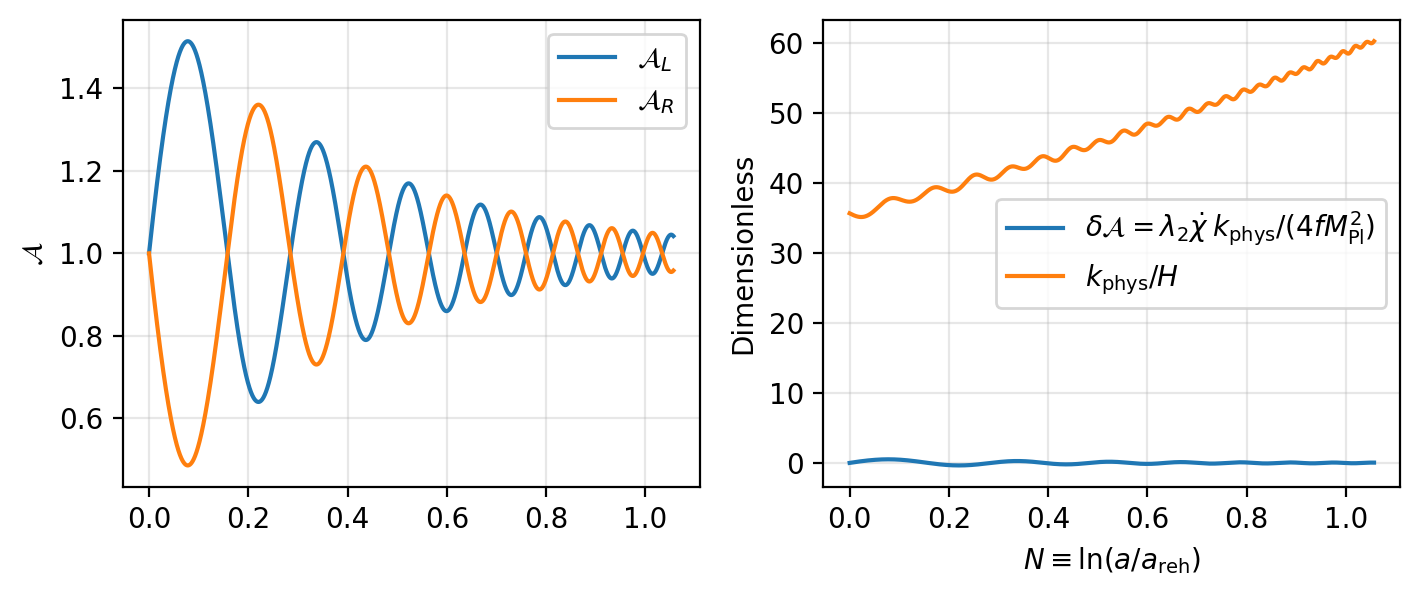}
\caption{Time dependence of the GCS kinetic coefficient during reheating for $\lambda_2=100$.
\textbf{Left:} ${\cal A}_{R,L}(x)$ for the two helicities (ghost-free requires ${\cal A}>0$).
\textbf{Right:} the parity-violating correction $\delta A$ and the ratio $k_{\rm phys}/H$ which grows as the universe expands.
}
\label{fig:calA}
\end{figure}

\begin{figure}[t]
\centering
\includegraphics[width=\linewidth]{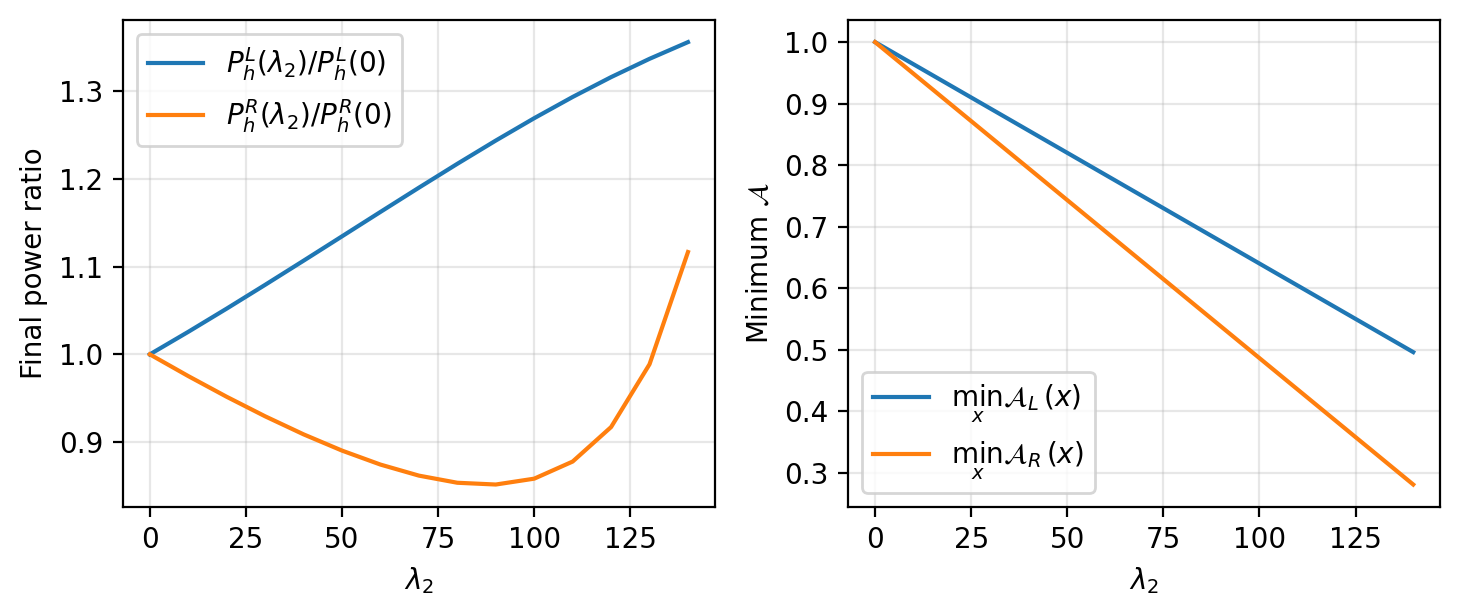}
\caption{Scan over $\lambda_2$ for the background.
\textbf{Left:} final tensor power ratios at $x_{\rm end}=0.5$.
\textbf{Right:} $\min_x{\cal A}_{R,L}(x)$, showing the approach toward the ghost boundary ${\cal A}=0$ as $\lambda_2$ increases.
}
\label{fig:scan}
\end{figure}

The system is oscillatory but does not exhibit strong stiffness for the parameters in Table~\ref{tab:params}. We verified numerically that shifting
both endpoints $(x_{\rm reh},x_{\rm end})\to (x_{\rm reh}+\Delta,x_{\rm end}+\Delta)$ leaves the final enhancement ratios
unchanged up to numerical roundoff ($\lesssim 10^{-11}$ for $\Delta=10$ and $\Delta=100$).

\begin{table}[t]
\centering
\begin{tabular}{c|c}
\hline
Parameter & Value (Figs.~\ref{fig:metric-vacuum} and \ref{fig:gauge-vacuum}) \\
\hline
$M_{\rm pl}$ & $1$ \\
$k$ & $1$ \\
$f$ & $1$ \\
$m_\varphi$ & $1$ \\
$m_\chi$ & $1$ \\
$g_A$ & $1$ \\
$\lambda_1$ (background) & $3$ \\
$\lambda_2$ (scan) & $5,\ 30,\ 70$ \\
\hline
Initial $x_{\rm reh}$ & $49$ \\
Final $x_{\rm end}$ & $0.25$ \\
Initial $a(x_{\rm reh})$ & $1$ \\
Initial $\varphi(x_{\rm reh})$ & $0.5$ \\
Initial $\pi_{\varphi}(x_{\rm reh})$ & $0$ \\
Initial $\Pi_\varphi(x_{\rm reh})$ & $0$ \\
Initial $\chi(x_{\rm reh})$ & $0.1$ \\
Initial $\pi_\chi(x_{\rm reh})$ & $0$ \\
Initial $\Pi_\chi(x_{\rm reh})$ & $0.05$ \\
Initial ${\cal Q}(x_{\rm reh})$ & $0.4$ \\
Initial $\pi_Q(x_{\rm reh})$ & $-0.4\mathcal{H}_{\text{reh}}$ \\
Initial $\Pi_{\cal Q}(x_{\rm reh})$ & $0$ \\
\hline
\end{tabular}
\caption{Parameter set used only for Figs.~\ref{fig:metric-vacuum} and \ref{fig:gauge-vacuum}, comparing the effect of $R\tilde{R}$ with and without the $F\tilde{F}$ coupling. Same notes from Table~\ref{tab:params} apply, except here we use a strongly-coupled regime.
These parameter values yield a visibly sourced gauge-vacuum contribution during reheating while maintaining a consistently expanding background.
}
\label{tab:params_fig8}
\end{table}

\begin{figure}[t]
\centering
\includegraphics[width=\linewidth]{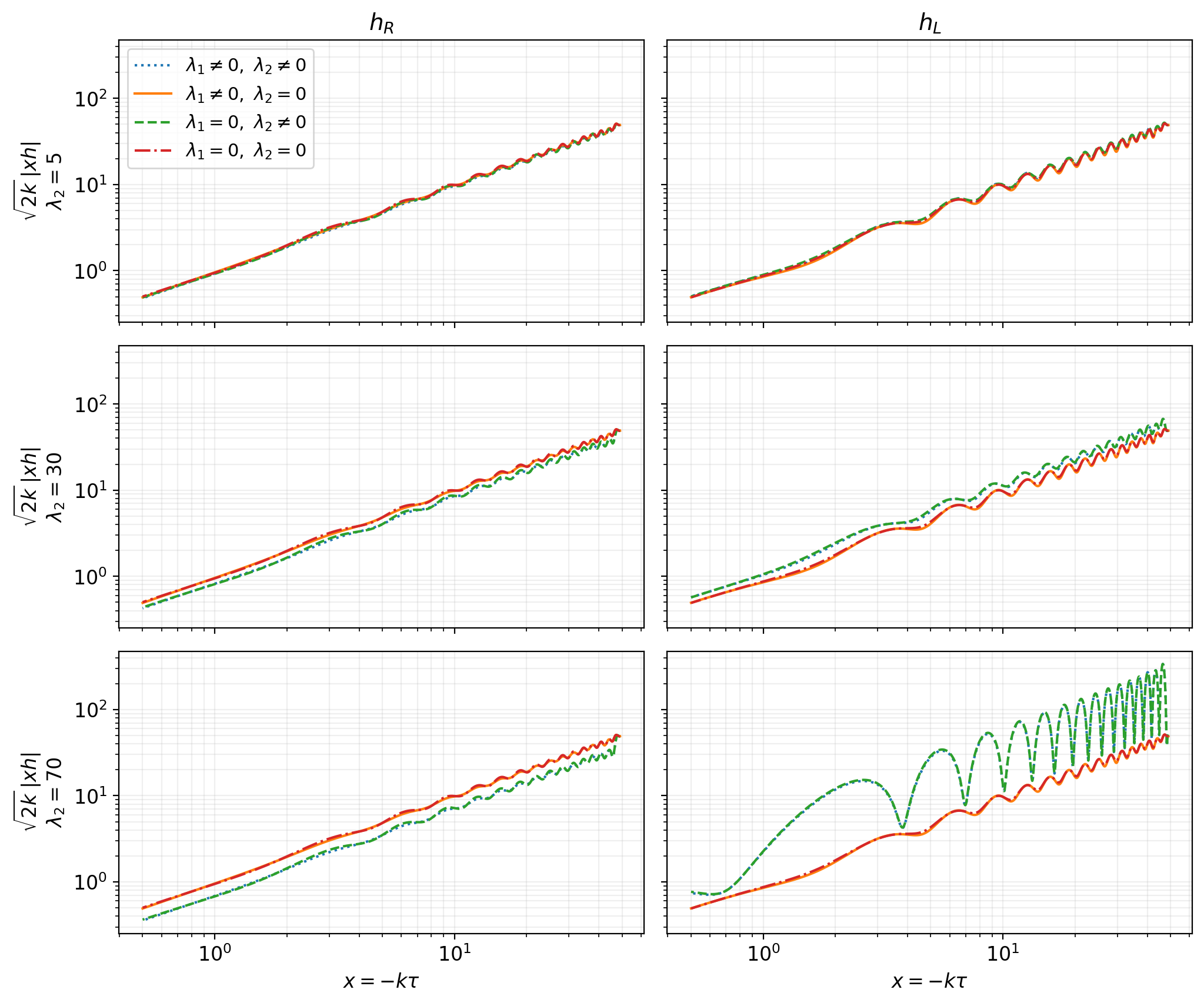}
\caption{\textbf{Metric-vacuum sector.}
We plot the canonically normalized metric tensor mode functions $\sqrt{2k}\,|x\,h_A|$ for $A=R$ (left column) and $A=L$ (right column),
with
(i) $\lambda_1\neq 0,\lambda_2\neq 0$ (both terms),
(ii) $\lambda_1\neq 0,\lambda_2=0$ (no GCS),
(iii) $\lambda_1=0,\lambda_2\neq 0$ (GCS only), and (iv) $\lambda_1=\lambda_2=0$ (neither term).
Rows correspond to $\lambda_2=\{5,30,70\}$.
During reheating, the oscillatory $\chi'(\tau)$ makes ${\cal A}(x)$ time-dependent, producing a helicity-dependent distortion even in the metric-vacuum sector. We emphasize that this figure corresponds to the more strongly-coupled benchmark (Table~\ref{tab:params_fig8}) than what we consider in the rest of this paper, to illustrate the effect of the $R\tilde{R}$ contribution.
}
\label{fig:metric-vacuum}
\end{figure}

\begin{figure}[t]
\centering
\includegraphics[width=\linewidth]{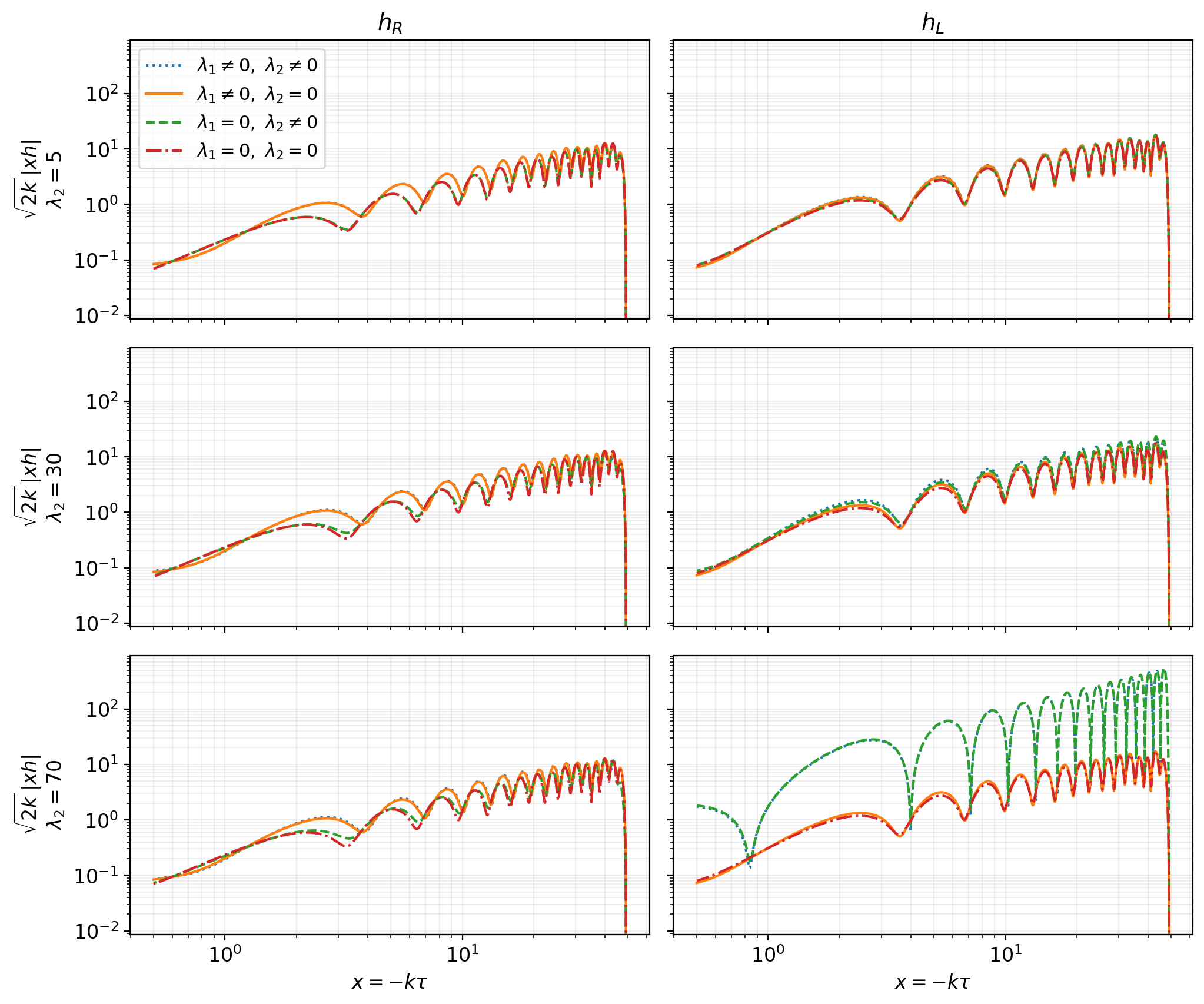}
\caption{\textbf{Gauge-vacuum sector.}
Same as Fig.~\ref{fig:metric-vacuum}, but for the gauge-vacuum initial conditions.
This sector captures the standard SU(2) sourcing contribution to $h_A$, while the GCS term changes the helicity-dependent propagation through ${\cal A}(x)$.  Similar to Fig.~\ref{fig:metric-vacuum}, we emphasize that this figure corresponds to the more strongly coupled benchmark (Table~\ref{tab:params_fig8}) than what we consider in the rest of this paper, to illustrate the effect of the $R\tilde{R}$ contribution.
}
\label{fig:gauge-vacuum}
\end{figure}

\section{Ghost-free condition} \label{sec:ghost-free}
During reheating, $\chi$ oscillates and so does $\dot\chi = \chi'/a$, making ${\cal A}(x)$ time-dependent.
Fig.~\ref{fig:calA} shows ${\cal A}_{R,L}$ for $\lambda_2=100$, together with the parity-violating correction
$\delta A\equiv \lambda_2\dot\chi\,k_{\rm phys}/(4fM_{\rm Pl}^2)$ and the growing ratio $k_{\rm phys}/H$. ${\cal A}(x)$ is always positive, so we do not have a ghost, but its deviation from unity is enough to affect the tensor perturbations.

Fig.~\ref{fig:scan} scans $\lambda_2$ and shows both the final tensor power ratios and the minimum value of ${\cal A}(x)$.
As expected, increasing $\lambda_2$ increases the size of the parity-violating effect, but eventually pushes ${\cal A}$ toward zero.

\section{$R\tilde{R}$ effect with and without $F\tilde{F}$}
The GCS term can enhance the helicity that is not directly sourced by the SU(2) tensor mode. To illustrate this, in Figs.~\ref{fig:metric-vacuum} and \ref{fig:gauge-vacuum} we plot the evolution of the tensor mode functions when the following terms are present: (i) both $\chi F\tilde F$ and $\chi R\tilde R$, (ii) $\chi F\tilde F$ only, (iii) $\chi R\tilde R$ only, and (iv) neither term.

Because turning off $\lambda_1$ would also change the background evolution, as a controlled diagnostic
we keep the reheating background fixed (computed with $\lambda_1\neq 0$), and we switch $\lambda_1$ and/or $\lambda_2$ only in the tensor-sector coefficients
${\cal S}$ and ${\cal C}$ (and ${\cal A}$ for $\lambda_2$)\footnote{We emphasize this is not a self-consistent theory truncation, but a diagnostic solely meant to demonstrate the $R\tilde{R}$ contribution with and without the $F\tilde{F}$ term.}.
This isolates the parity-violating propagation effect of $\chi R\tilde R$ from the usual SU(2) sourcing effect, while retaining the reheating-specific feature:
$\chi'(\tau)$ oscillates, so ${\cal A}(x)$ is time-dependent.

For visibility we use a second, more strongly coupled reheating benchmark (Table~\ref{tab:params_fig8}),
chosen so that the gauge-vacuum contribution is not negligible.
In Figs.~\ref{fig:metric-vacuum} and \ref{fig:gauge-vacuum}, we plot the tensor mode evolution for the metric-vacuum and gauge-vacuum sectors, respectively.
We show three representative values of $\lambda_2=\{5,30,70\}$, which are still ghost-free during the reheating interval of interest.

\bibliographystyle{JHEP}
\bibliography{bib}

\end{document}